\documentclass[aps,pre,preprint,superscriptaddress,showpacs,amssymb,floatfix]{revtex4}
\usepackage{graphicx}
\usepackage{graphics}
\usepackage{amsmath}
\usepackage{tabularx}
\usepackage{color}
\usepackage{doi}
\usepackage{textcomp}
\newcolumntype{Y}{>{\centering\arraybackslash}X}

\begin{document}

\title{Droplet Finite-Size Scaling of the Majority Vote Model on Quenched Scale-Free Networks}

\author{D. S. M. Alencar}
\affiliation{Departamento de Física, Universidade Federal do Piauí, 57072-970, Teresina - PI, Brazil}
\author{T. F. A. Alves}
\affiliation{Departamento de Física, Universidade Federal do Piauí, 57072-970, Teresina - PI, Brazil}
\author{F. W. S. Lima}
\affiliation{Departamento de Física, Universidade Federal do Piauí, 57072-970, Teresina - PI, Brazil}
\author{R. S. Ferreira}
\affiliation{Departamento de Ciências Exatas e Aplicadas, Universidade Federal de Ouro Preto, 35931-008, João Monlevade - MG, Brazil}
\author{G. A. Alves}
\affiliation{Departamento de Física, Universidade Estadual do Piauí, 64002-150, Teresina - PI, Brazil}
\author{A. Macedo-Filho}
\affiliation{Departamento de Física, Universidade Estadual do Piauí, 64002-150, Teresina - PI, Brazil}

\date{Received: date / Revised version: date}

\begin{abstract}

We consider the Majority Vote model coupled with scale-free networks. Recent works point to a non-universal behavior of the Majority Vote model, where the critical exponents depend on the connectivity while the network's effective dimension $D_\mathrm{eff}$ is unity for a degree distribution exponent $5/2<\gamma<7/2$. We present a finite-size theory of the Majority Vote Model for uncorrelated networks and present generalized scaling relations with good agreement with Monte-Carlo simulation results. The presented finite-size theory has two main sources of size dependence. The first source is an external field describing a mass media influence on the consensus formation and the second source is the scale-free network cutoff. The model indeed presents non-universal critical behavior where the critical exponents depend on the degree distribution exponent $5/2<\gamma<7/2$. For $\gamma \geq 7/2$, the model is on the same universality class of the Majority Vote model on Erdös-Renyi random graphs, while for $\gamma=7/2$, the critical behavior presents additional logarithmic corrections.

\end{abstract}


\maketitle

\section{Introduction}

We consider a consensus formation model \cite{Galam-2008}, called Majority Vote (MV) model \cite{Oliveira-1992, Pereira-2005, Yu-2017, Wu-2009, Crochik-2005, Vilela-2009, Lima-2012, Vieira-2016, Krawiecki-2018, Stanley-2018, Alves-2019} on uncorrelated scale-free networks. We are interested in the finite-size scaling behavior of the MV model on scale-free networks, which presents a rich feature where the power-law degree fluctuations can change the expected mean-field behavior. The scale-free property induces a non-universal behavior, depending on the degree distribution exponent \cite{Krawiecki-2019}.

Ref.\ \cite{Krawiecki-2019} presents a heterogeneous mean-field (HMF) theory of the MV model on unbounded (i.e., with no limit on the number of hub connections) annealed scale-free networks. However, it is known that the scale-free property also induces network correlations \cite{Cohen-2010, Barabasi-2016}. A fundamental feature of a scale-free network is the presence of hubs, which are the highly connected nodes \cite{Cohen-2010, Barabasi-2016} which are responsible for the ultra small-world property, and the presence of degree correlations. In addition, if hubs are present, we can expect a change in the system behavior \cite{Cohen-2010, Barabasi-2016}.

Results of Ref.\ \cite{Krawiecki-2019} are consistent with a non-universal critical behavior for $5/2 < \gamma < 7/2$. In the case of $\gamma>7/2$, we have the same universality class of the MV model on random Erdös-Renyi graphs, where $\gamma$ is the degree distribution exponent. The same non-universal behavior is reported in Ref.\ \cite{Lima-2006}, on MV model on Barabasi-Albert (BA) networks with two opinion states, while maintaining the effective dimension $D_{\mathrm{eff}}$, defined as
\begin{equation}
D_\mathrm{eff} \equiv 2\beta/\nu + \gamma'/\nu,
\label{effectivedimension}
\end{equation}
equal to unity, where $\beta$, $\gamma'$, and $\nu$ are the order parameter, susceptibility, and shifting exponents, respectively. In addition, Ref.\ \cite{Vilela-2020} considered a modified version of the MV model on BA networks where the individuals can have three discrete opinions and its results pointed to varying $1/\nu$, $\beta/\nu$ and $\gamma/\nu$ exponent ratios when changing $z$, also reporting $D_\mathrm{eff}=1$.

Unbounded degree fluctuations introduce non-trivial effects on phase transitions \cite{Castellano-2006, Hong-2007, Dorogovtsev-2008}. One well-studied example is the Contact Process (CP) model on the Uncorrelated Configuration Model (UCM) \cite{Catanzaro-2005, Dorogovtsev-2008, Hong-2007}. The UCM is an algorithm to generate uncorrelated scale-free networks with an externally controlled power-law exponent $\lambda$ \cite{Catanzaro-2005}, which we can impose a structural cutoff (maximum number of hub connections) to generate uncorrelated scale-free networks.

The critical behavior of the CP model on UCM networks was subjected by an intense debate if the CP model obeys the Heterogeneous Mean Field (HMF) theory for unbounded scale-free networks \cite{Castellano-2006, Hong-2007, Ferreira-2011b}, settled by the fact that the critical behavior of the CP model on scale-free networks is subjected to finite-size scaling corrections, depending on the degree distribution cutoff. Considering the special case of UCM networks with $\gamma=3$, HMF theory predicts logarithmic corrections to scaling \cite{Ferreira-2011b}. In the same way, results from a special Mean-Field theory, applied to BA networks, predict an extra logarithmic dependence in the critical behavior of the CP model order parameter \cite{Odor-2012}. 

In this way, we present a theory for the finite-size corrections on the MV model scaling on the uncorrelated scale-free networks. The main sources of the scale-free corrections are tracked to be the network cutoff, which is required to build an uncorrelated scale-free network and a droplet external field that induces a small variation of the magnetization, scaling as $N^{-1}$. We compared our approach with simulation results on UCM and BA networks \cite{Barabasi-1999, Albert-2002, Newman-2002, Dorogovtsev-2003, Boccaletti-2006, Barrat-2008, Cohen-2010, Barabasi-2016}.

This paper is organized as follows: in Sec.\ \ref{hmfsection}, we extend results of Ref.\ \cite{Krawiecki-2019}, in Sec.\ \ref{dropletsection}, we describe the droplet finite-size scaling relations for the MV model, in Sec.\ \ref{simulationsection}, we discuss our simulation results, and in Sec.\ \ref{conclusions}, we present our final considerations.

\section{Revisiting the Heterogeneous Mean-Field Theory\label{hmfsection}}

\subsection*{Dynamics of MV model}

In this work, we consider the two-state MV model \cite{Oliveira-1992, Pereira-2005, Yu-2017}, which can describe a ferromagnetic material in contact with two heat baths, one at zero temperature and the other at infinite temperature. Also, the MV model can describe consensus formation whose dynamics has the following rules:
\begin{enumerate}
   \item We consider a network with $N$ nodes. We assign a system state
   \begin{equation}
     \boldsymbol{\sigma} = \left( \sigma_1, \sigma_2, ..., \sigma_N \right),
   \end{equation}
   where each network node is associated to a stochastic variable $\sigma_{i} = \pm 1$, corresponding to two opinion states for each network node. We can start the dynamics by randomly selecting the opinion state for each node;
   \item At each time step, we randomly choose one node $i$ to be updated;
   \item Then, we try a spin flip with rate $w_i$, written as
\begin{equation}
   w_i(\boldsymbol{\sigma}) = \frac{\alpha}{2} \left[1-(1-2q)\sigma_{i}S\left(\sum_{\delta=1}^{k_i} \sigma_{\delta} \right)\right],
\label{mvfliprate}
\end{equation}   
where the index summation $\delta$ runs over the $k_i$ nearest neighbors of the node $i$, $\alpha=1$ is a parameter with an inverse time dimension, related to the selection probability, and $S(x)$ is the \textit{signal} function, which summarizes the neighborhood majority opinion
\begin{equation}
S(x) = \begin{cases}
          -1, & \text{if } x < 0; \\
          0,  & \text{if } x = 0; \\
          1,  & \text{if } x > 0. \\
       \end{cases}
\label{signalfunction}
\end{equation}
\end{enumerate}

The noise parameter $q$ in Eq.\ \ref{mvfliprate} induces a continuous phase transition from a consensus phase to a no-consensus phase, analogous to the ferro-paramagnetic phase transition. In the context of consensus formation models of sociophysics, the noise parameter is a social temperature that gives the probability of local contrarians. Eq.\ \ref{mvfliprate} summarizes a Markovian process where an individual will oppose its neighborhood opinion with probability $q$ and follows its neighborhood with probability $1-q$. In case of no local majority, the node $j$ can assume any opinion state with $w_i=1/2$.

The MV model dynamics obeys the following Master Equation \cite{Tome-2015}, valid for local spin-flips
\begin{equation}
\frac{d}{dt} \mathcal{P}_{\boldsymbol{\sigma}} = \sum_i^N w_i(\boldsymbol{\sigma^i}) \mathcal{P}_{\boldsymbol{\sigma^i}} - w_i(\boldsymbol{\sigma}) \mathcal{P}_{\boldsymbol{\sigma}},
\label{masterequation}
\end{equation}
where $\mathcal{P}_{\boldsymbol{\sigma}}$ is the occupation probability of one system state, and
\begin{equation}
  \boldsymbol{\sigma^i} = \left( \sigma_1, \sigma_2, ...,-\sigma_i, ..., \sigma_N \right),
\end{equation}
is the system state after a well succeed spin-flip from the system state $\boldsymbol{\sigma}$. An average ensemble of the local magnetization for local spin-flip dynamics should give
\begin{equation}
  \frac{\partial}{\partial t} \left< \sigma_i \right> = - 2\left< \sigma_iw_i(\boldsymbol{\sigma}) \right>.
  \label{langevin}
\end{equation}

\subsection*{Time evolution of the order parameter for unbounded networks}

From Eqs.\ \ref{mvfliprate}, and \ref{langevin}, we can obtain the mass-action equation of the local magnetization $\left< \sigma_i \right>$
\begin{equation}
  \frac{\partial}{\partial t} \left< \sigma_i \right> = - \left< \sigma_i \right> + \lambda\left< S\left(\sum_{\delta=1}^{k_j} \sigma_{\delta} \right) \right>,
  \label{mvmassaction}
\end{equation}
where we defined 
\begin{equation}
  \lambda \equiv 1-2q.
  \label{noise}
\end{equation}
Following Ref.\ \cite{Krawiecki-2019}, we can write
\begin{equation}
  \left< S\left(\sum_{\delta=1}^{k_j} \sigma_{\delta} \right) \right> = \sum_{\sigma_i=\pm1} \sigma_i P^{\text{majority}}_i(\sigma_i),
 \label{meanneighborconsensus}
\end{equation}
where $P^{\text{majority}}_i$ is the majority opinion distribution of the $k_i$ neighbors of node $i$. The majority opinion distribution $P^{\text{majority}}_i$ of the $k_i$ neighbors of node $i$ in Eq.\ \ref{meanneighborconsensus} is given by
\begin{equation}
P^{\text{majority}}_i(\sigma) = \sum_{\ell=\lceil k_i/2 \rceil}^{k_i} \binom{k_i}{\ell} \prod_j^{\ell} P^{\text{node}}_j(\sigma)\prod_{j'}^{k_i-\ell} P^{\text{node}}_{j'}(-\sigma),
\label{majoritydistribution}
\end{equation}
where $\lceil x \rceil$ is the ceiling function, and the indices $j,j'$ run on the neighborhood of node i. We note that any event with $\lceil k_i/2 \rceil < \ell < k_i$ neighbors aligned with $\sigma_i$ and $k_i-\ell$ neighbors with $-\sigma_i$ contributes to $P^{\text{majority}}_i(\sigma_i)$. In addition, $P^{\text{node}}_i(\sigma)$ is the distribution of local values of $\sigma = \pm 1$, which depends on the local magnetization. In annealed networks, local properties should depend only on the node degree, therefore
\begin{equation}
  P^{\text{node}}_j(\sigma) = P^{\text{node}}_k(\sigma) = \frac{1 + \sigma\left< \sigma_k \right>}{2},
\end{equation}
where $\left< \sigma_k \right>$ is the local magnetization of a node $j$ with degree $k$. Note that if the local magnetization vanishes, we should have $P^{\text{node}}_j(\sigma) = 1/2$.

We now consider uncorrelated networks, where each factor in Eq.\ \ref{majoritydistribution} would have the same frequency, given by
\begin{equation}
  P\left( k \mid k' \right) = \frac{k'P(k')}{\left<k\right>},
  \label{conditionalprobability}
\end{equation}
which is the conditional probability that a neighbor of a node with degree $k$ should have degree $k'$. The conditional probability of uncorrelated networks $P\left( k \mid k' \right)$ should be proportional to the network degree distribution $P(k)$ and the neighbor degree $k'$. Therefore, we can rewrite Eq.\ \ref{majoritydistribution} for uncorrelated networks as
\begin{equation}
P^{\text{majority}}_k(\sigma) = \sum_{\ell=\lceil k/2 \rceil}^{k} \binom{k}{\ell} \prod_{k'}^{k} \left( \frac{1 + \sigma\left< \sigma_k \right>}{2} \right)^{\frac{k'P(k')}{\left<k\right>}\ell} \left( \frac{1 - \sigma\left< \sigma_k \right>}{2} \right)^{\frac{k'P(k')}{\left<k\right>}(k-\ell)}.
\label{neutralmajoritydistribution}
\end{equation}
In addition, by using the binomial expansion until linear terms, we can further simplify Eq.\ \ref{neutralmajoritydistribution} as
\begin{equation}
P^{\text{majority}}_k(\sigma) \approx \sum_{\ell=\lceil k/2 \rceil}^{k} \binom{k}{\ell} \prod_{k'}^{k} \left( \frac{1 + \sigma M'}{2} \right)^{\ell} \left( \frac{1 - \sigma M'}{2} \right)^{k-\ell},
\label{approxmajoritydistribution}
\end{equation}
valid in the $\left< \sigma_k \right> << 1$ regime, where we defined the rescaled order parameter
\begin{equation}
  M' \equiv \frac{1}{\left< k \right>} \sum_k kP(k)\left< \sigma_k \right>,
  \label{rescaledmagnetization}
\end{equation}
where every node is weighted by its degree. The central limit theorem can be used to give a continuous degree approximation for $P^{\text{majority}}_k(\sigma)$
\begin{equation}
  P^{\text{majority}}_k(\sigma) \approx \sqrt{\frac{2}{k\pi}}\int_{k/2}^{k} \exp{\left\lbrace -\left[ t - \frac{k}{2}\left(1+\sigma M'\right)\right]^2\right\rbrace}dt,
  \label{continuousmajoritydistribution}
\end{equation}
which can be approximated for well-connected networks, where hubs satisfy $k \gg 1$, to the simpler expression
\begin{equation}
  P^{\text{majority}}_k(\sigma) \approx \frac{1}{2} + \frac{\sigma}{2}\text{erf}\left( \sqrt{\frac{k}{2}} M' \right),
  \label{finalmajoritydistribution}
\end{equation}
where $\text{erf}(x)$ is the error function, defined as
\begin{equation}
  \text{erf}(x) \equiv \frac{2}{\sqrt{\pi}}\int_{0}^{x} \exp{\left(-t^2\right)}dt.
\end{equation}

Returning to Eq.\ \ref{meanneighborconsensus}, and substituting $P^{\text{majority}}_k(\sigma)$ given in Eq.\ \ref{finalmajoritydistribution}, we obtain
\begin{equation}
  \left< S\left(\sum_{\delta=1}^{k_j} \sigma_{\delta} \right) \right> \approx \text{erf}\left( \sqrt{\frac{k}{2}} M' \right),
 \label{approxneighborconsensus}
\end{equation}
which depends on the rescaled magnetization, written in Eq.\ \ref{rescaledmagnetization}. From Eq.\ \ref{approxneighborconsensus}, we can 
recast Eq.\ \ref{mvmassaction} as
\begin{equation}
  \frac{\partial}{\partial t} \left<\sigma_k\right> = - \left<\sigma_k\right> + \lambda\ \text{erf}\left( \sqrt{\frac{k}{2}} M' \right),
  \label{mvmassaction2}
\end{equation}
and multiplying Eq.\ \ref{mvmassaction2} with $P\left( k \mid k' \right)$ written in Eq.\ \ref{conditionalprobability} and summing in $k$, we obtain an evolution equation for $M'$, depending on the neighboring consensus average written in Eq.\ \ref{approxneighborconsensus}.
\begin{equation}
  \frac{\partial}{\partial t}M' = -M' + \lambda \sum_k \frac{kP(k)}{\left<k\right>} \text{erf}\left( \sqrt{\frac{k}{2}} M' \right).
  \label{rescaledmagnetizationdynamics}
\end{equation}

In the case of unbounded power-law networks with the normalized degree distribution, where we do not limit the number of connections of the hubs, we have the following distribution
\begin{equation}
P(k) = (\gamma-1)m^{\gamma-1}k^{-\gamma},
\label{degreedistribution}
\end{equation}
where $m$ is the minimum number of connections, we can write the stationary rescaled magnetization $M'$ in the continuous degree limit
\begin{equation}
  M' = \lambda(\gamma-2)m^{\gamma-2}\int_m^\infty k^{-\gamma+1} \text{erf}\left( \sqrt{\frac{k}{2}} M'\right)dk,
  \label{stationaryrescaledmagnetization1}
\end{equation}
from Eq.\ \ref{rescaledmagnetizationdynamics}. We can integrate the right side of Eq.\ \ref{stationaryrescaledmagnetization1} by parts to write
\begin{equation}
  M' = \lambda\ \text{erf}\left( \sqrt{\frac{m}{2}} M'\right) + \lambda \sqrt{\frac{m}{2\pi}}M'\left(\frac{mM'^2}{2}\right)^{\gamma-5/2}\Gamma\left(-\gamma+5/2,\frac{mM'^2}{2}\right),
  \label{stationaryrescaledmagnetization}
\end{equation}
where $\Gamma(s,x)$ is an incomplete Gamma function. Eq.\ \ref{stationaryrescaledmagnetization} is a transcendent recursive equation for $M'$, which can only be solved in the linear asymptotic limit $M'\to 0$.

Note that we considered uncorrelated networks, while we do not impose any cutoff on the number of hub connections. Hubs can induce degree correlations, and the unbounded number of connections can turn the network into a disassortative one. We have to impose a cutoff in the distribution to preserve the neutral feature of the network. The cutoff is also a source of finite-size corrections, as we analyze in Sec.\ \ref{dropletsection}.

\subsection*{Asymptotic expression of the order parameter for unbounded networks}

We can use the error function expansion
\begin{equation}
  \text{erf}(x) = \frac{2}{\sqrt{\pi}} \left( x - \frac{x^3}{3} + \frac{x^5}{10} - \cdots \right),
  \label{errorexpansion}
\end{equation}
and the asymptotic expansion of the incomplete Gamma function for $x \to 0$
\begin{equation}
  x^{s}\Gamma\left(-s,x\right) = \frac{\pi}{\sin\left[\pi(s+1)\right]} \frac{x^s}{\Gamma(s+1)}
  + \frac{1}{s} + \frac{x}{1-s} + \frac{x^2}{2\left(s-2\right)} + \cdots,
  \label{gammaexpansion}
\end{equation}
for a non-integer $s$ to write the asymptotic behavior of $M'$ for $M'\to 0$
\begin{eqnarray}
  && M'\left[ 2\lambda\sqrt{\frac{m}{2\pi}}\frac{\gamma-2}{\gamma-5/2} - 1 - \frac{2\lambda}{3}\sqrt{\frac{m}{2\pi}}\frac{\gamma-2}{\gamma-7/2}\frac{mM'^2}{2} + \right. \nonumber \\
  && \qquad \left. - \lambda\sqrt{\frac{m}{2\pi}}\frac{\pi}{\Gamma(\gamma-3/2)\left|\sin[\pi(\gamma-3/2)]\right|}\left(\frac{mM'^2}{2}\right)^{\gamma-5/2}\right] \approx 0
  \label{assymptoticrescaledmagnetization}
\end{eqnarray}
valid for $\gamma \ne 5/2$ and $\gamma \ne 7/2$. We shall return to the case $\gamma=7/2$. We can solve for $M'$ in Eq.\ \ref{assymptoticrescaledmagnetization}, where we can identify two cases:
\begin{enumerate}
  \item[a)] For $5/2 < \gamma < 7/2$, we can neglect the third term in Eq.\ \ref{assymptoticrescaledmagnetization}. Solving for $M'$ yields two roots given by
  \begin{equation}
  (M'^2)^{\gamma-5/2} = \begin{cases}
              \frac{2\lambda\sqrt{\frac{m}{2\pi}}\frac{\gamma-2}{\gamma-5/2}-1}
              {\lambda\sqrt{\frac{m}{2\pi}} \frac{\pi}{\Gamma(\gamma-3/2)\left|\sin[\pi(\gamma-3/2)]\right|} 
              \left(\frac{m}{2}\right)^{\gamma-5/2}},                                & \quad \text{if } \lambda >    \lambda_c; \\
              0,                                                                     & \quad \text{if } \lambda \leq \lambda_c; \\
           \end{cases}
  \label{assymptoticsforgammacase2}
  \end{equation}
  where we obtain $\beta = 1/[2(\gamma-5/2)]$.
  \item[b)] For $\gamma > 7/2$, we can neglect the last term between brackets in Eq.\ \ref{assymptoticrescaledmagnetization}. Again, solving for $M'$ yields two roots, now given by
  \begin{equation}
    M'^2 = \begin{cases}
              \frac{2\lambda\sqrt{\frac{m}{2\pi}}\frac{\gamma-2}{\gamma-5/2}-1}
              {\frac{2\lambda}{3}\sqrt{\frac{m}{2\pi}}\frac{\gamma-2}{\gamma-7/2}\frac{m}{2}},  & \quad \text{if } \lambda >     \lambda_c; \\
              0,                                                                                & \quad \text{if } \lambda \leq  \lambda_c; \\
           \end{cases}
  \label{assymptoticsforgammacase1}
  \end{equation}
  where we readily obtain the critical order parameter exponent $\beta = 1/2$.

\end{enumerate}
In both cases, we identify a critical threshold that separates the paramagnetic phase with $M'=0$, and the ferromagnetic phase with $M' \ne 0$, given by
\begin{equation}
  \lambda_c = \frac{1}{2}\sqrt{\frac{2\pi}{m}}\frac{\gamma-5/2}{\gamma-2},
  \label{criticalthreshold}
\end{equation}
which reproduces the same result of Ref.\ \cite{Krawiecki-2019}. From the expression of the critical threshold in Eq.\ \ref{criticalthreshold}, we can conclude that the model presents a vanishing threshold for $\gamma = 5/2$.

Now, we return to the case $\gamma=7/2$. We can combine Eq.\ \ref{stationaryrescaledmagnetization} with $\gamma=7/2$, Eq.\ \ref{errorexpansion}, and the following expansion
\begin{equation}
  x\Gamma(-1,x) = 1 + x\left( \ln x + \gamma_{\text{em}} + 1 \right) - \frac{x^2}{2} + \cdots,
  \label{gammaexpansionlog}
\end{equation}
where $\gamma_{\text{em}}$ is the Euler-Mascheroni constant, to obtain
\begin{equation}
  M'\left[3\lambda\sqrt{\frac{m}{2\pi}} - 1 + \frac{1}{2}\frac{1}{\sqrt{2\pi}}m^{3/2}M'^2\left| \ln M' \right| \right] \approx 0.
  \label{assymptoticrescaledmagnetizationlog}
\end{equation}
and solving for $M'$ yields
  \begin{equation}
  M'^2 \left| \ln M'\right| = \begin{cases}
              \frac{3\lambda\sqrt{\frac{m}{2\pi}}-1}
              {\frac{\lambda}{\sqrt{\pi}}\left(\frac{m}{2}\right)^{3/2}}, & \quad \text{if } \lambda >     \lambda_c; \\
              0,                                                          & \quad \text{if } \lambda \leq  \lambda_c; \\
           \end{cases}
  \label{assymptoticsforgammacase3}
  \end{equation}
in a way we obtain $\beta=1/2$ with additional logaritmic corrections. The critical threshold $\lambda_c$ is also given by Eq.\ \ref{criticalthreshold} with $\gamma=7/2$.

In summary, we extended the results of Ref.\ \cite{Krawiecki-2019} to include the asymptotic relations of the rescaled magnetization expressed in Eqs.\ \ref{assymptoticsforgammacase1}, \ref{assymptoticsforgammacase2}, and \ref{assymptoticsforgammacase3}, for the cases $5/2 < \gamma < 7/2$, $\gamma > 7/2$, and $\gamma=7/2$, respectively. In addition, we found the explicit form of logarithmic corrections at $\gamma = 7/2$, proportional to $\left| \ln M'\right|$. We also reproduced the result for the critical exponent $\beta$ of Ref.\ \cite{Krawiecki-2019}, given by
\begin{equation}
  \beta = \begin{cases}
          \frac{1}{2(\gamma-5/2)}, & \quad \text{if } 5/2 < \gamma < 7/2; \\
          \frac{1}{2}            , & \quad \text{if } \gamma \geq 7/2
  \end{cases}
\end{equation}
where the case $\gamma = 7/2$ presents additional logarithmic corrections, and for $\gamma = 5/2$, we have a vanishing threshold.

The results of this session are applied to unbounded networks without the scale-free property. A scale-free network should have a cutoff in the degree distribution to maintain its neutral, uncorrelated nature. In general, in uncorrelated networks, there are other sources of scaling corrections, as seen for the SIR model \cite{Satorras-2001, Moreno-2002, Boguna-2013}, and the contact process \cite{Castellano-2006, Castellano-2008, Boguna-2009, Ferreira-2011a, Ferreira-2011b, Odor-2012}, coming from the network cutoff. In the next session, we present a theory for finite-size scaling corrections on scale-free networks.

\section{Droplet finite-size scaling for the MV model\label{dropletsection}}

\subsection*{Cutoff power-law networks}

We now consider power-law networks with the following distribution
\begin{equation}
P(k) = \begin{cases}
          \frac{\gamma-1}{f(\gamma-1)}m^{\gamma-1}k^{-\gamma},  & \quad \text{if } m \leq k \leq k_c; \\
          0,                                                    & \quad \text{if } k < m \text{ and if } k > k_c; \\
     \end{cases}
  \label{cutoffdegreedistribution}
\end{equation}
where $f(x)$ is written as
\begin{equation}
  f(x) = 1 - \left( \frac{m}{k_c} \right)^x,
  \label{correction-f}
\end{equation}
which expresses the effect of a cutoff $k_c$, and in general, we have the cutoff in the form
\begin{equation}
  k_c = N^{1/\omega},
  \label{cutoff}
\end{equation}
where $\omega=2$ in the case of the structural cutoff imposed on UCM networks \cite{Catanzaro-2005}, and $\omega = \gamma-1$ in the case of a natural cutoff seen in growing network models as the BA model \cite{Barabasi-2016}. The degree moments of the distribution in Eq.\ \ref{cutoffdegreedistribution} are given by
\begin{equation}
  \langle k^{\ell} \rangle = \frac{(\gamma -1)}{(\gamma-\ell-1)}\frac{f(\gamma-\ell-1)}{f(\gamma-1)}m^{\ell},
  \label{degree-moments}
\end{equation}
where $f(x)$, and $k_c$ are given by Eqs.\ \ref{correction-f}, and \ref{cutoff}, respectively. 

\subsection*{Dynamic evolution with an external field}

In order to link the dynamic behavior of the system with the finite size of the underlying network, we use the external field. The external field generally interacts with the individual spins by a Zeeman interaction. However, in consensus formation models, the external field describes the mass media's influence over the individuals, by favoring one of the possible opinion states. We modify the spin-flip rate in Eq.\ \ref{mvfliprate} to
\begin{equation}
  w_j(\boldsymbol{\sigma}) = \frac{\alpha}{2} \left(1 - p_h\right) \left[1-(1-2q)\sigma_{j}S\left(\sum_{\delta=1}^{k_j} \sigma_{\delta} \right)\right] + \alpha p_h,
  \label{fieldmvrate}
\end{equation}
which can be interpreted as the spin trying an independent spin-flip with probability $p_h$ given by
\begin{equation}
  p_h = \frac{h}{2}\left(1 - \sigma_i\right),
\end{equation}
before the usual MV spin-flip, where we note that $p_h = h$ is the rate that the spins with $\sigma_i = -1$ will flip while spins with $\sigma_i = 1$ would have $p_h = 0$. The external field breaks $\mathbb{Z}^2$ symmetry by favoring the $\sigma_i = 1$ state.

In the case of a small external field that produces a droplet small variation of the total magnetization that scales as
\begin{equation}
h\Delta M \sim \frac{2}{N},
  \label{dropletscaling}
\end{equation}
corresponding to only one spin flip in a network with $N$ nodes, we can write an approximate spin-flip rate from Eq.\ \ref{fieldmvrate}, given by
\begin{equation}
  w_j(\boldsymbol{\sigma}) \approx \frac{\alpha}{2} \left[1-(1-2q)\sigma_{j}S\left(\sum_{\delta=1}^{k_j} \sigma_{\delta} \right)\right] + \alpha p_h,
  \label{fieldmvrateapprox}
\end{equation}
and from the Master Equation, we obtain the analogous of Eq.\ \ref{mvmassaction} in the presence of a small external field
\begin{equation}
    \frac{\partial}{\partial t} \left< \sigma_i \right> = - \left< \sigma_i \right> + \lambda\left< S\left(\sum_{\delta=1}^{k_j} \sigma_{\delta} \right) \right> + h,
    \label{fieldmvmassaction}
\end{equation}
where we neglected an additive term $h\left< \sigma_i \right>$ which scales as $2/N$, $\lambda$ is given as a function of the noise in Eq.\ \ref{noise}, and $S(x)$ is defined in Eq.\ \ref{signalfunction}. From Eq.\ \ref{approxneighborconsensus}, we can obtain the evolution of the local magnetization 
\begin{equation}
  \frac{\partial}{\partial t} \left<\sigma_k\right> = - \left<\sigma_k\right> + \lambda\ \text{erf}\left( \sqrt{\frac{k}{2}} M' \right) + h,
  \label{fieldmvmassaction2}
\end{equation}
and the evolution of the rescaled magnetization in the presence of an external field
\begin{equation}
  \frac{\partial}{\partial t}M' = -M' + \lambda \sum_k \frac{kP(k)}{\left<k\right>} \text{erf}\left( \sqrt{\frac{k}{2}} M' \right) + h.
  \label{fieldrescaledmagnetizationdynamics}
\end{equation}

\subsection*{Asymptotic expressions in an external field}

The stationary solution for the rescaled order parameter with the external field is readily obtained in an analogous way to the previous section and is given in the continuous limit as
\begin{equation}
  M' = \lambda \frac{\gamma-2}{f(\gamma-2)} \int_m^{k_c} k^{-\gamma+1} \text{erf}\left( \sqrt{\frac{k}{2}} M' \right)dk + h,
  \label{fieldrescaledmagnetization}
\end{equation}
where we used the cutoff power-law distribution written in Eq.\ \ref{cutoffdegreedistribution}. Integrating into the right side and applying expansions in Eqs.\ \ref{errorexpansion}, and \ref{gammaexpansion} to Eq.\ \ref{fieldrescaledmagnetization}, we obtain for the asymptotic behavior of the rescaled magnetization $M'$ in the $M' \to 0$ limit
\begin{equation}
  h \approx \left( 1 - \frac{\lambda}{\lambda_c} \right)M' + \frac{\lambda}{6} \sqrt{\frac{2}{\pi}} gM'^3 + \mathcal{O}(M'^5),
  \label{fieldassymptoticrescaledmagnetization}
\end{equation}
where the critical threshold $\lambda_c$ is
\begin{equation}
  \lambda_c = \sqrt{\frac{\pi}{2}}\frac{\left<k\right>}{\left<k^{3/2}\right>} = \frac{1}{2}\sqrt{\frac{2\pi}{m}}\frac{\gamma-5/2}{\gamma-2}\frac{f(\gamma-2)}{f(\gamma-5/2)},
  \label{finitecriticalthreshold}
\end{equation}
and the correction factor $g$ is given by
\begin{equation}
  g = \frac{\left<k^{5/2}\right>}{\left<k\right>} = \frac{\gamma-2}{\gamma-7/2}\frac{f(\gamma-7/2)}{f(\gamma-2)}m^{3/2}
  \label{correctionfactor1}
\end{equation}
From Eq.\ \ref{correction-f}, we see that $f(x) \to 1$ for $x>0$, in a way that the critical threshold in Eq.\ \ref{finitecriticalthreshold} reproduces the expression in Eq.\ \ref{criticalthreshold} for $k_c \to \infty$.

We can also obtain an asymptotic expression of the magnetization $M$ as a function of the rescaled magnetization $M'$ as follows. The finite-size scaling of the magnetization can be determined from Eq.\ \ref{fieldassymptoticrescaledmagnetization} if we express it as a function of the rescaled magnetization $M'$. The magnetization $M$ is defined as
\begin{equation}
  M \equiv \sum_k P(k) \left< \sigma_k \right>.
  \label{magnetization}
\end{equation}
From Eq.\ \ref{fieldmvmassaction2}, we can obtain the stationary state of $\left< \sigma_k \right>$
\begin{equation}
  \left< \sigma_k \right> = \lambda\ \text{erf} \left( \sqrt{\frac{k}{2}} M' \right) + h,
\end{equation}
and substitution of $\left< \sigma_k \right>$ in Eq.\ \ref{magnetization} yields in the continuous limit
\begin{equation}
  M - h = \lambda\ \frac{\gamma-1}{f(\gamma-1)}m^{\gamma-1}\int_m^{k_c}k^{-\gamma} \text{erf} \left( \sqrt{\frac{k}{2}} M' \right) dk,
\end{equation}
where we substituted the cutoff power-law distribution written in Eq.\ \ref{cutoffdegreedistribution}. In the same way, we obtained Eq.\ \ref{fieldassymptoticrescaledmagnetization}, we can write
\begin{equation}
  M - h \approx \lambda \sqrt{\frac{2}{\pi}} aM' - \mathcal{O}(M'^3),
  \label{magnetizationfromrescaledmagnetization}
\end{equation}
where we defined
\begin{equation}
  a = \left< k^{1/2} \right> = \frac{\gamma-1}{\gamma-3/2}\frac{f(\gamma-3/2)}{f(\gamma-1)} m^{1/2}.
  \label{correctionfactor2}
\end{equation}
We note that $M$ and $M'$ will have the same scale in the thermodynamic limit for $h \to 0$.

\subsection*{Finite-size scaling on an uncorrelated scale-free network}
 
From the asymptotic expressions, we can obtain the finite-size scaling for small magnetizations $M$, when close to the critical threshold. We start from Eq.\ \ref{fieldassymptoticrescaledmagnetization} for $\lambda=\lambda_c$, combined with Eq.\ \ref{magnetizationfromrescaledmagnetization} with $h \to 0$, which yields
\begin{equation}
  h \propto \frac{g}{a^3}M^3,
\end{equation}
and integrating in $M$, we obtain
\begin{equation}
  h \Delta M \propto \frac{g}{a^3}M^4,
\end{equation}
and by using the droplet scaling of $h\Delta M$ at Eq.\ \ref{dropletscaling}, we obtain
\begin{equation}
  M \propto \left(\frac{gN}{a^3}\right)^{-1/4}.
  \label{magnetizationscaling}
\end{equation}

We can also obtain the shifting scaling from Eq.\ \ref{fieldassymptoticrescaledmagnetization} for $h=0$, which yields
\begin{equation}
  \frac{\lambda}{\lambda_c} - 1 \propto gM'^2,
  \label{shiftingscaling1}
\end{equation}
and substituting Eqs.\ \ref{magnetizationfromrescaledmagnetization}, and \ref{magnetizationscaling} in \ref{shiftingscaling1}, we obtain
\begin{equation}
  \frac{\lambda}{\lambda_c} - 1 \propto \left(\frac{aN}{g}\right)^{-1/2}.
  \label{shiftingscaling}
\end{equation}

We can summarize results in Eqs.\ \ref{magnetizationscaling}, and \ref{shiftingscaling} in the following scaling form
\begin{equation}
  M = \left( \frac{gN}{a^3} \right)^{-1/4} F\left[ \left( \frac{aN}{g} \right)^{1/2} \left( \frac{\lambda}{\lambda_c} - 1 \right) \right],
  \label{scalingform}
\end{equation}
where $g$ and $a$ are given in Eqs.\ \ref{correctionfactor1}, and \ref{correctionfactor2}, respectively. The correction factor $a$ does not change the system scaling in the thermodynamic limit, while $g$ can change the critical exponents in the $5/2 < \gamma < 7/2$ interval. The correction factor $g$ obeys the following mesoscopic scaling
\begin{equation}
  g \sim \begin{cases}
  \frac{\gamma-2}{7/2-\gamma}\frac{f(7/2-\gamma)}{f(\gamma-2)}m^{\gamma-2}N^{(7/2-\gamma)/\omega}, & \quad \text{if } 5/2 < \gamma < 7/2; \\
  \frac{3/2}{f(3/2)}m^{3/2}\left|\ln\left( \frac{m}{N^{1/\omega}} \right)\right|,                  & \quad \text{if } \gamma=7/2; \\
  \frac{\gamma-2}{\gamma-7/2}\frac{f(\gamma-7/2)}{f(\gamma-2)}m^{3/2},                             & \quad \text{if } \gamma>7/2;
     \end{cases}
  \label{mesoscopiccorrectionfactorscaling}
\end{equation}
where we substituted the explicit cutoff dependence at Eq.\ \ref{cutoff} in Eq.\ \ref{correctionfactor1}. 

The next section compares the theoretical results for the finite-size critical behavior with simulation results for the kinetic dynamics on quenched uncorrelated networks. The Droplet theory is exact for annealed networks. However, as we will show, the droplet theory presented in this section also works on quenched uncorrelated networks obeying the degree distribution at Eq.\ \ref{cutoffdegreedistribution}.

\section{Simulation results\label{simulationsection}}

\subsection*{Observables and critical behavior}

We present the needed observables to investigate the critical behavior of the MV model in the following. The main observable is the opinion balance $o$, analogous to the magnetization of magnetic equilibrium systems
\begin{equation}
o = \left\vert \frac{1}{N} \sum_i s_i \right\vert.
\end{equation}
From the opinion balance, one can calculate the order parameter by averaging $m$. In quenched networks, we should do a quenched average, done on random realizations of the network. For each random realization, one should evolve dynamics to a stationary state, and then collect an ensemble composed of a time series. The order parameter $M$, the susceptibility $\chi$, and Binder's fourth-order cumulant $U$ are given by the following relations, respectively \cite{Oliveira-1992}
\begin{eqnarray}
M    &=& \left[ \langle o \rangle \right], \nonumber \\
\chi &=& \left[ N (\langle o^{2} \rangle - \langle o \rangle ^{2}) \right], \nonumber \\
U    &=& \left[ 1 - \frac{\langle o^{4} \rangle}{3\langle o^{2} \rangle^{2}} \right], 
\label{observables}
\end{eqnarray}
where the symbol $\langle ... \rangle$ represents the average of a time series and the symbol $\left[ ... \right]$ represents the quench average. All observables are functions of $\lambda$. 

From the results of the previous section, notably Eq.\ \ref{scalingform}, we conjecture that the observables written in Eq.\ \ref{observables} should obey the following finite-size scaling (FSS) relations
\begin{eqnarray}
M    &=&  \left( \frac{gN}{a^3} \right)^{-1/4} F_M \left[ \left( \frac{aN}{g} \right)^{1/2} \left( \frac{\lambda}{\lambda_c} - 1 \right) \right], \nonumber \\
\chi &=&  \left( \frac{N}{ga^3} \right)^{1/2} F_\chi \left[ \left( \frac{aN}{g} \right)^{1/2} \left( \frac{\lambda}{\lambda_c} - 1 \right) \right], \nonumber \\
U    &=& F_U \left[ \left( \frac{aN}{g} \right)^{1/2} \left( \frac{\lambda}{\lambda_c} - 1 \right) \right].
\label{observables-fss}
\end{eqnarray}
where $M$ now indicates an average on the stationary state and on network realizations, differently of Sec.\ \ref{dropletsection} where $M$ indicated a stationary state of the magnetization of an annealed network.

To obtain the relevant observables, we performed Monte Carlo Markov chains (MCMC) on UCM networks with different sizes $N$. We simulated $160$ random network realizations for each size to make quench averages. For each network replica, we considered $10^5$ MCMC steps to let the system evolve to a stationary state and another $10^5$ MCMC steps to collect $10^5$ values of the opinion balance to measure the observables. One MCMC step for the MV model is defined as the update of $N$ spins. Error bars were calculated by resampling data \cite{Landau-2015}.

\subsection*{Results and Discussion}

\begin{figure}[ht]
\begin{center}
\includegraphics[scale=0.12]{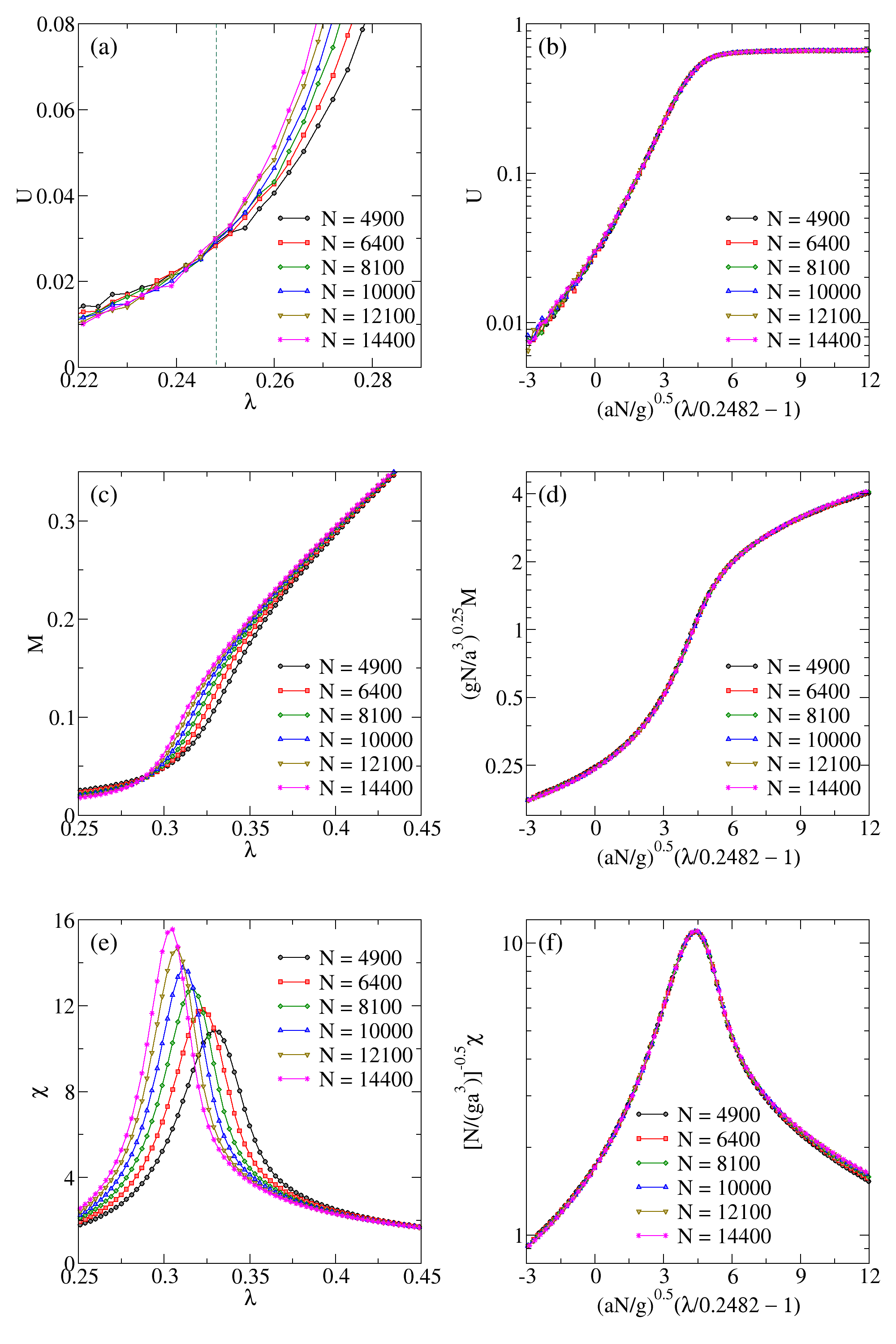}
\end{center}
\caption{(Color Online) We show the stationary simulation results of averages in Eq.\ \ref{observables}, in (a), (c), and (e). respectively, for the Majority Vote coupled to UCM networks with $\gamma=3.0$ and $m=8$, as functions of the control parameter $\lambda$. We show data collapses of Binder cumulant $U$, magnetization $M$, and Susceptibility $\chi$, in (b), (d), and (f), respectively, according to Eq.\ \ref{observables-fss}. The critical threshold is estimated as $\lambda_c=0.2482$. For $5/2 < \gamma < 7/2$, the critical behavior is non-universal, where the critical exponent ratios depend on the degree exponent $\gamma$.}
\label{mv-collapsed-results-3.0}
\end{figure}

We show our simulation results for the MV model on UCM networks with $\gamma=3.0$ and $m=8$ in Fig.\ \ref{mv-collapsed-results-3.0}. We show data collapses by using the scaling relations written on Eq.\ \ref{observables-fss}. All simulation results are consistent with a continuous phase transition with critical thresholds $\lambda_c$ depending on the degree distribution exponent $\gamma$. In addition, we obtained a good agreement between simulation results and the scaling relation predicted by droplet theory in the case of magnetization and Binder cumulant. In the case of the susceptibility, we \textit{ad hoc} propose a scaling form with a good agreement with simulation data.

\begin{figure}[ht]
\begin{center}
\includegraphics[scale=0.12]{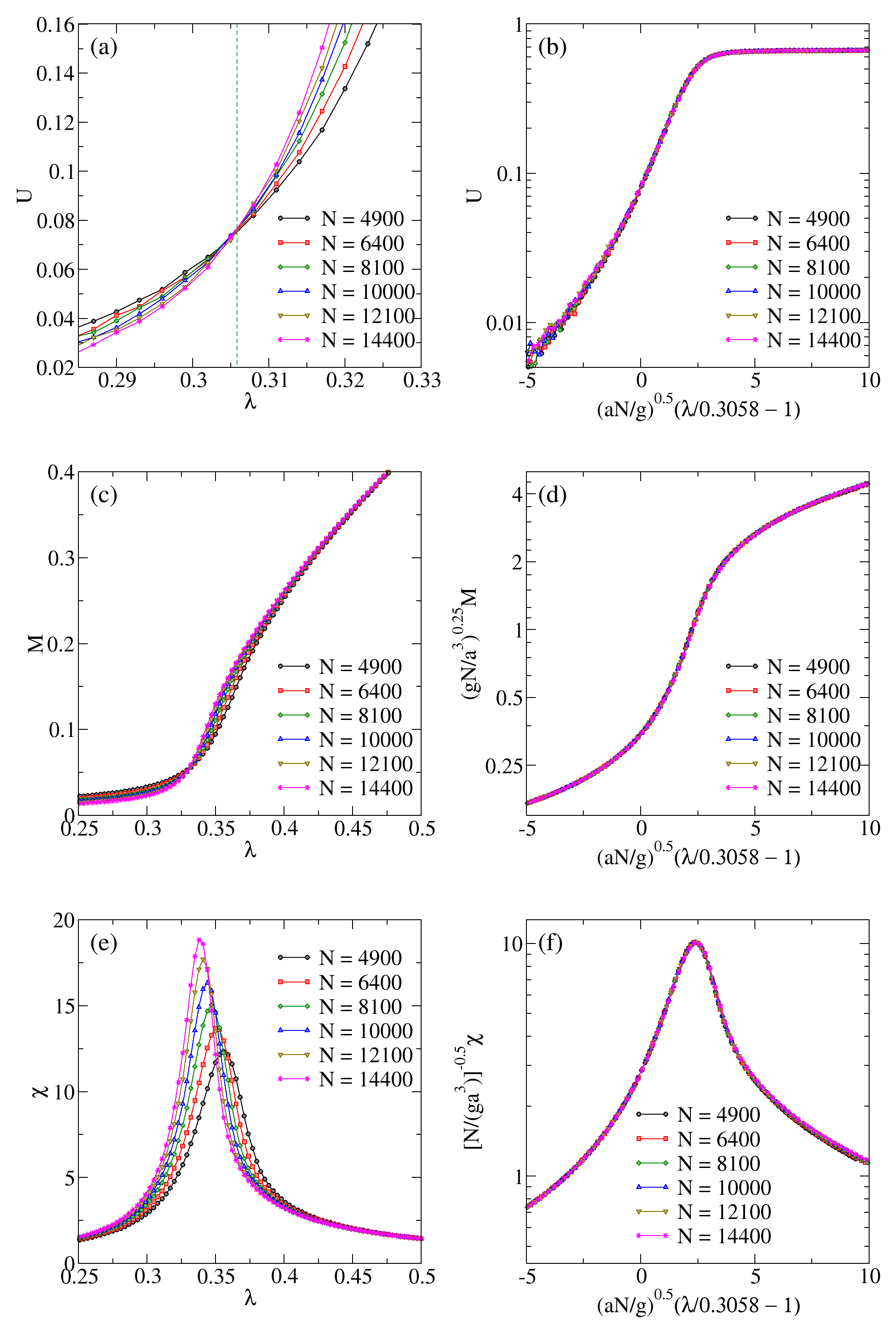}
\end{center}
\caption{(Color Online) The same of Fig.\ \ref{mv-collapsed-results-3.0} for $\gamma=3.5$. The critical threshold is estimated as $\lambda_c=0.3058$. For $\gamma = 7/2$, we have the same universality class of the MV model on random Erdös-Renyi graphs \cite{Pereira-2005}, where the scaling presents additional logarithmic corrections.}
\label{mv-collapsed-results-3.5}
\end{figure}

\begin{figure}[ht]
\begin{center}
\includegraphics[scale=0.12]{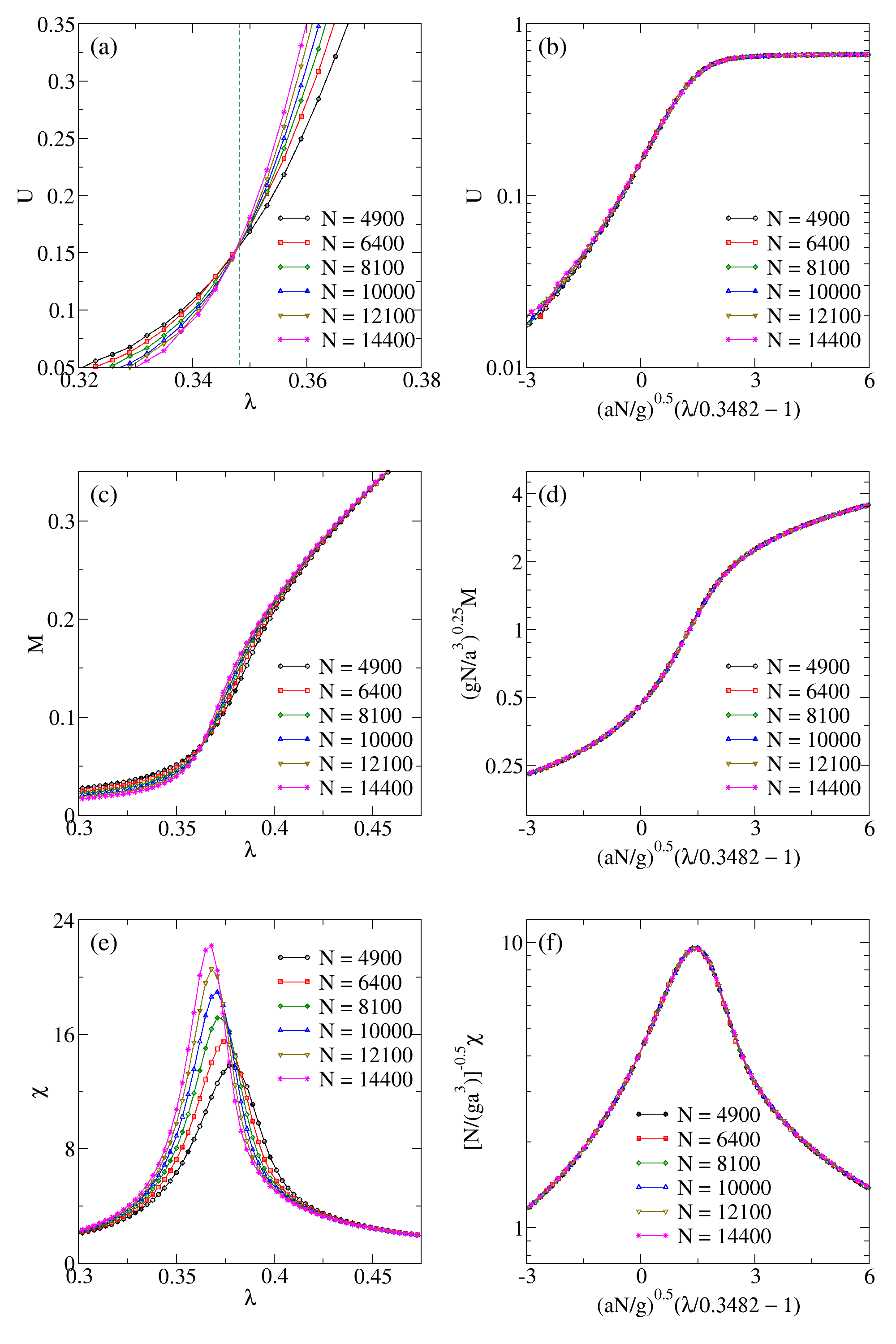}
\end{center}
\caption{(Color Online) The same of Fig.\ \ref{mv-collapsed-results-3.0} for $\gamma=4.0$. The critical threshold is estimated as $\lambda_c=0.3482$. For $\gamma \geq 7/2$, the universality class of the model is the same as the MV model on random Erdös-Renyi graphs \cite{Pereira-2005}.}
\label{mv-collapsed-results-4.0}
\end{figure}

We also simulated results on UCM networks with $\gamma=3.25$ (not shown), $\gamma=3.5$ (Fig.\ \ref{mv-collapsed-results-3.5}), $\gamma=3.75$ (not shown), and $\gamma=4$ (Fig.\ \ref{mv-collapsed-results-4.0}). We obtained a dependence on $\gamma$ exponent for quenched UCM networks, contrary to our prediction in Eq.\ \ref{finitecriticalthreshold}. However, the data collapse presents a very good agreement with scaling relations in Eq.\ \ref{observables-fss}. In addition, we tested the scaling relations on Eq.\ \ref{observables-fss} to the BA networks (not shown) and obtained good collapses, as a consequence that BA networks are almost uncorrelated \cite{Dorogovtsev-2008}.

The critical behavior of the MV model on uncorrelated networks depends on the correction factor $g$. From the mesoscopic scale of $g$ in Eq.\ \ref{mesoscopiccorrectionfactorscaling}, we can obtain the asymptotic scale of $g$ in the thermodynamic limit ($N \to \infty$)
\begin{equation}
  g \sim \begin{cases}
          N^{\frac{\gamma-7/2}{\omega}}, & \quad \text{if } 5/2 < \gamma < 7/2; \\
          \ln N,                         & \quad \text{if } \gamma = 7/2; \\   
          \text{constant},               & \quad \text{if } \gamma > 7/2
  \end{cases}
  \label{correctionfactorscaling}
\end{equation}
and from the scaling relations in Eq.\ \ref{observables-fss}, we can obtain the critical exponent ratios by substituting the $g$ scaling. Recalling the fact that the magnetization should vanish at the critical threshold as $N^{-\beta/\nu}$, we can write for the critical exponent ratio $\beta/\nu$
\begin{equation}
  \frac{\beta}{\nu} = \begin{cases}
          \frac{1}{4} + \frac{\gamma-7/2}{4\omega}, & \quad \text{if } 5/2 < \gamma < 7/2; \\
          \frac{1}{4},                              & \quad \text{if } \gamma \geq 7/2;
  \end{cases}
    \label{betaovernu}
\end{equation}
and also, the shifting of the susceptibility maxima from the critical threshold should scale as $N^{-1/\nu}$. We obtain the critical exponent ratio $1/\nu$
\begin{equation}
  \frac{1}{\nu} = \begin{cases}
          \frac{1}{2} - \frac{\gamma-7/2}{2\omega}, & \quad \text{if } 5/2 < \gamma < 7/2; \\
          \frac{1}{2},                              & \quad \text{if } \gamma \geq 7/2;
  \end{cases}
  \label{1overnu}
\end{equation}
and the fact that the susceptibility should diverge as $N^{\gamma'/\nu}$ yields the critical exponent ratio $\gamma'/\nu$
\begin{equation}
  \frac{\gamma'}{\nu} = \begin{cases}
          \frac{1}{2} - \frac{\gamma-7/2}{2\omega}, & \quad \text{if } 5/2 < \gamma < 7/2; \\
          \frac{1}{2},                              & \quad \text{if } \gamma \geq 7/2
  \end{cases}
  \label{gammaovernu}
\end{equation}

The critical dimension, defined on Eq.\ \ref{effectivedimension}, from Eqs.\ \ref{betaovernu}, and \ref{gammaovernu}, is unity, which is consistent with the results of Refs.\ \cite{Lima-2006, Krawiecki-2019}. We also note that the finite result by the inclusion of the network cutoff should result in a different expected critical behavior from the heterogeneous Mean-Field theory presented in Sec.\ \ref{hmfsection}, as a consequence of the network cutoff, needed to maintain the uncorrelated nature of the scale-free networks. 

Finally, we summarize the critical behavior of the MV model on uncorrelated networks: For $5/2 < \gamma < 7/2$, the critical behavior is non-universal, where the critical exponent ratios depend on the degree exponent $\gamma$ as seen in Eqs.\ \ref{betaovernu}, \ref{1overnu}, and \ref{gammaovernu}. For $\gamma \geq 7/2$, we have the same universality class of the MV model on random Erdös-Renyi graphs \cite{Pereira-2005}, where the scaling presents additional logarithmic corrections in the case $\gamma=7/2$.

\section{Conclusions\label{conclusions}}

We considered a consensus formation model, namely the MV model on quenched networks. Our simulation results suggest a continuous phase transition, where the critical noises depend on network connectivity and the degree exponent $\gamma$. We also presented a finite-size scaling theory where the correction factors on finite networks should change the critical behavior of the model. For $5/2 < \gamma < 7/2$, we obtained a non-universal critical behavior, where the critical exponent ratios depend on the degree exponent $\gamma$. In the case of $\gamma>7/2$, we have the same universality class of the MV model on random Erdös-Renyi graphs. For $\gamma=7/2$, the critical behavior of the system is also the same as the MV model on random Erdös-Renyi graphs, however, with additional logarithmic corrections.

\section{Acknowledgments}

We would like to thank CAPES (Coordenação de Aperfeiçoamento de Pessoal de Nível Superior), CNPq (Conselho Nacional de Desenvolvimento Científico e tecnológico), FUNCAP (Fundação Cearense de Apoio ao Desenvolvimento Científico e Tecnológico) and FAPEPI (Fundação de Amparo a Pesquisa do Estado do Piauí) for the financial support. We acknowledge using Dietrich Stauffer Computational Physics Lab, Teresina, Brazil, and Laboratório de Física Teórica e Modelagem Computacional - LFTMC, Teresina, Brazil, where the simulations were performed.

\bibliography{textv1}

\end{document}